# A New Way to See Inside Black Holes

Richard Conn Henry[1], James Overduin[1,2] and Kielan Wilcomb[2]

[1]Department of Physics and Astronomy
Johns Hopkins University
3400 North Charles Street
Baltimore, MD, 21218, USA

[2]Department of Physics, Astronomy and Geosciences
Towson University
8000 York Road
Towson, MD, 21252, USA
E-mail: joverduin@towson.edu

## Abstract

Black holes are real astrophysical objects, but their interiors are hidden and can only be "observed" through mathematics. The structure of rotating black holes is typically illustrated with the help of special coordinates. But any such coordinate choice necessarily results in a distorted view, just as the choice of projection distorts a map of the Earth. The truest way to depict the properties of a black hole is through quantities that are *coordinate-invariant*. We compute and plot all the independent curvature invariants of rotating, charged black holes for the first time, revealing a landscape that is much more beautiful and complex than usually thought.

Black holes are one of the most spectacular predictions of Einstein's theory of General Relativity: regions of spacetime so strongly curved by gravity that not even light can escape. They are known to exist as the end products of the collapse of very massive stars, and at the cores of most galaxies. But they are not accessible to observation in the same way as other phenomena in nature. They are surrounded by event horizons, surfaces from which no information can emerge. Thus black holes are objects for which the usual course of scientific discovery is reversed. Usually we observe first, and use mathematics and art later on to organize and explain what we have observed. With black holes, it is just the opposite: we are guided by mathematics and art, and we hope to observe later!

Perhaps for that reason, popular depictions of these objects are often misleading. Typically they are shown as two-dimensional "whirlpools in space", as in the 1979 Disney film *The Black Hole*. Spatially they are of course three-dimensional, resembling a "hole from every direction," as memorably explained in the 2014 film *Interstellar*. Their effects on space and time outside the horizon can be suggested by the use of colored grid lines (Figure 1, left). Here the change in color and radial compression of the grid lines represent the warping of time and space respectively due to the mass of the black hole, while the spiral shape of the grid lines in the horizontal plane suggests the twisting of spacetime due to the hole's spin. These effects are known as gravitational redshift, geodetic effect and frame-dragging respectively, and all have recently been detected experimentally in the weak gravitational field of the Earth [1]. Because of mathematical analogies with Maxwell's equations for electromagnetism, the warping effects are sometimes referred to as "gravito-electric", while the twisting is "gravito-magnetic".

Fortunately, we have mathematical solutions for all possible types of black holes: those with no charge or spin (Schwarzschild), those with spin but no charge (Kerr), those with charge but no spin (Reissner-Nordstrom) and, most generally of all, those with both charge and spin (Kerr-Newman). Most astrophysical black holes are likely to be of the Kerr type since charge will gradually be neutralized through the preferential accumulation of oppositely charged matter. The interior of a Kerr black hole is typically illustrated using Boyer-Lindquist coordinates, in which the radius of the event horizon is $r = m + \sqrt{m^2 - a^2}$ where *m* and *a* are the black hole's mass and angular momentum (or spin). In such coordinates the event horizon appears as a spherical shell (inner surface in Figure 1, right).

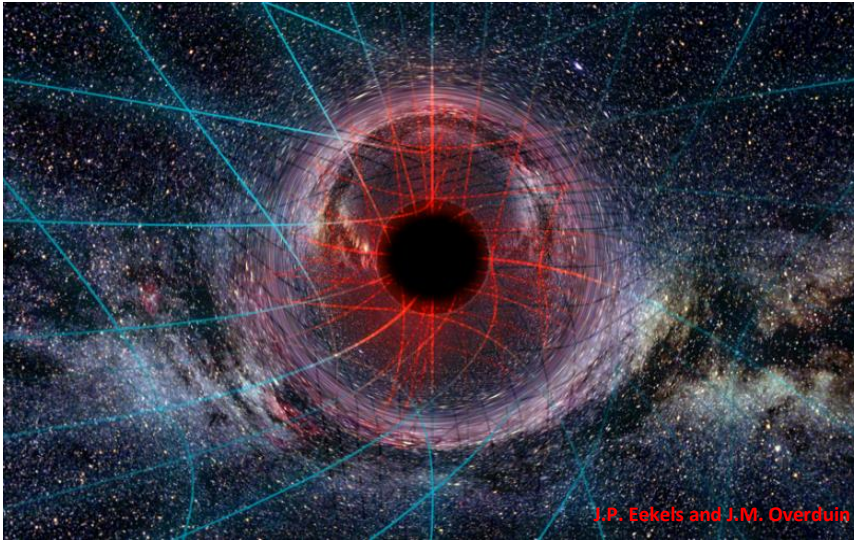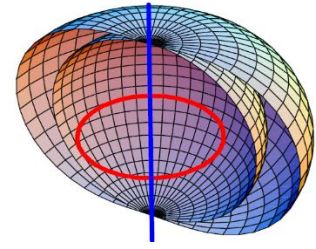

**Figure 1**: *Effects of a spinning black hole on space and time outside the horizon (left), and interior of a spinning black hole in Boyer-Lindquist coordinates (right), showing the ring singularity and event horizon (spherical inner shell).*

Such a picture has its uses, but is also misleading in important ways. For example, it suggests that spatial curvature everywhere on the horizon is *constant* and *positive*. This is incorrect, in exactly the same way that it is incorrect to say that Greenland is larger than the continental U.S.A., even though it may look that way on a flat (Mercator projection) map of the Earth. The error in both cases is an artifact of the choice of coordinates. The only way to illustrate the interior of a black hole without such distortion is to plot only *invariants*; that is, quantities whose value is the same regardless of the coordinates used.

We wish to obtain invariants characterizing the curvature of spacetime for the most general possible (Kerr-Newman) case. The essence of General Relativity is that the curvature so described is identical with what we feel as gravity. The starting point is the metric tensor $g_{ij}$, a generalization of the Pythagorean Theorem $d^2 = x^2 + y^2$ in which increments of distance in curved four-dimensional spacetime obey $ds^2 = \sum_{i,j} g_{ij} dx^i dx^j$ for coordinates $x^i$ (indices $i, j, \ldots$ range from 1 to 4). Taking $x^i$ to consist of time *t* plus the Boyer-Lindquist radius *r* and spherical polar coordinates $\theta, \varphi$, the Kerr-Newman metric is specified by $ds^2 = -\frac{\Delta}{\rho^2}(dt - a\sin^2\theta\, d\varphi)^2 + \frac{\sin^2\theta}{\rho^2}[(r^2 + a^2)d\varphi - a\, dt]^2 + \frac{\rho^2}{\Delta} dr^2 + \rho^2 d\theta^2$ where $\Delta \equiv r^2 - 2mr + a^2 + q^2$, $\rho^2 \equiv r^2 + a^2 \cos^2\theta$, and *q* is the black hole's electric charge [2].

Curvature is described in terms of the metric by the Riemann tensor $R^i{}_{jkl} \equiv \frac{\partial}{\partial x^k}\Gamma^i{}_{jl} - \frac{\partial}{\partial x^l}\Gamma^i{}_{jk} + \Gamma^m{}_{jl}\Gamma^i{}_{mk} - \Gamma^m{}_{jk}\Gamma^i{}_{ml}$ where $\Gamma^i{}_{jk} \equiv \frac{1}{2}g^{li}\left(\frac{\partial g_{kl}}{\partial x^j} + \frac{\partial g_{jl}}{\partial x^k} - \frac{\partial g_{kj}}{\partial x^l}\right)$ is the "connection". (These expressions assume the Einstein summation convention; i.e., summation over repeated indices.) But while the elements of $R^i{}_{jkl}$ describe the curvature, they do not do so in an *invariant* manner, since they are functions of the chosen coordinates. To describe the curvature in a coordinate-invariant way (what Einstein called "expressing thoughts without words"), one must construct scalar quantities from the metric and Riemann tensor. In general, there are fourteen of these [3], or as many as seventeen when allowance is made for certain degenerate cases [4]. It can be shown that only two of these (labelled $I_1$ and $I_2$ in [4]) are independent for the case of neutral black holes. Charged black holes are characterized by three more independent curvature scalars ($I_6$, $I_9$ and $I_{10}$). So one needs five such quantities to fully characterize the curvature of spacetime inside all possible time-independent black holes.

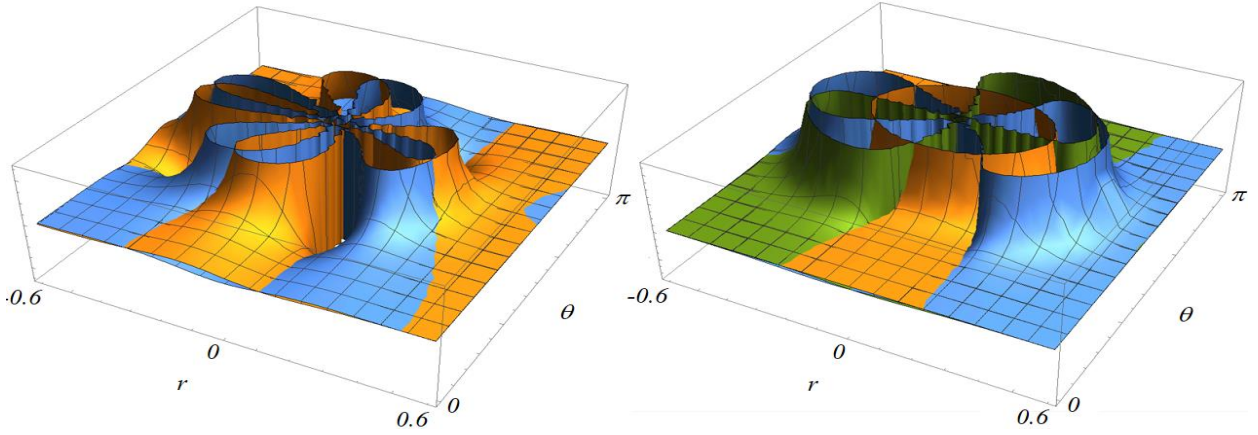

**Figure 2**: *The Weyl curvature invariants $I_1$ and $I_2$ (yellow and blue; left) and the Ricci and mixed curvature invariants $I_6$, $I_9$ and $I_{10}$ (yellow, blue and green; right), plotted as functions of $r$ and $\theta$ for a black hole of mass m=1, charge q=0.8 and spin a=0.6.*

Two of these curvature invariants have been previously discussed for the case $q = 0$ [5,6]. We have calculated and plotted all seventeen invariants in the general case $q \neq 0$ for the first time. Space constraints prevent us from listing them all here, but the five key independent quantities are defined by $I_1 = C_{ij}{}^{kl}C_{kl}{}^{ij}$ and $I_2 \equiv -C_{ij}{}^{kl}C^*{}_{kl}{}^{ij}$ (the *Weyl invariants*), $I_6 \equiv R_{ij}R^{ij}$ (the *Ricci invariant*), and $I_9 = C_{ikl}{}^{j}R^{kl}R_j{}^i$ and $I_{10} = -C^*{}_{ikl}{}^{j}R^{kl}R_j{}^i$ (the *mixed invariants*). Here $C_{ijkl} \equiv R_{ijkl} + \frac{R}{6}(g_{ik}g_{jl} - g_{il}g_{jk}) - \frac{1}{2}(g_{ik}R_{jl} - g_{il}R_{jk} - g_{jk}R_{il} + g_{jl}R_{ik})$ is the Weyl tensor, whose dual is $C^*{}_{ijkl} \equiv \frac{1}{2}\varepsilon_{ijmn}C^{mn}{}_{kl}$ where $\varepsilon^{ijkl}$ is the Levi-Civita tensor density ($\equiv +1$ when *ijkl* is an even permutation of 1234; $-1$ if an odd permutation; and 0 otherwise). The Ricci tensor and curvature scalar are defined by $R_{ij} = R^k{}_{ikj}$ and $R = R^i_i$ respectively. Indices are raised and lowered in these expressions by means of the metric tensor and its inverse, so that (for instance) the Weyl tensor in mixed form is $C_{ij}{}^{kl} = g^{km}g^{ln}C_{ijmn}$.

We have evaluated these quantities by extending a publicly available suite of symbolic *Mathematica* codes [7]. The results are plotted in Figure 2 above; mathematically, they read:

$$I_1 = \frac{8}{(r^2 + a^2\cos^2\theta)^6}[6m^2(r^6 - 15r^4a^2\cos^2\theta + 15r^2a^4\cos^4\theta - a^6\cos^6\theta)$$
$$- 12mq^2r(r^4 - 10r^2a^2\cos^2\theta + 5a^4\cos^4\theta) + q^4(6r^4 - 36r^2a^2\cos^2\theta + 6a^4\cos^4\theta)],$$

$$I_2 = \frac{96a\cos\theta}{(r^2 + a^2\cos^2\theta)^6}[m^2r(3r^4 - 10r^2a^2\cos^2\theta + 3a^4\cos^4\theta) - mq^2(5r^4 - 10r^2a^2\cos^2\theta + a^4\cos^4\theta)$$
$$+ 2q^4r(r^2 - a^2\cos^2\theta)],$$

$$I_6 = \frac{4q^4}{(r^2 + a^2\cos^2\theta)^4},$$

$$I_9 = \frac{16q^4[r^2(q^2 - mr) - a^2(q^2 - 3mr)\cos^2\theta]}{(r^2 + a^2\cos^2\theta)^7} \quad \text{and} \quad I_{10} = \frac{16aq^4\cos\theta[r(3mr - 2q^2) - a^2m\cos^2\theta]}{(r^2 + a^2\cos^2\theta)^7}.$$

The plots in Figure 2 may be likened to topographical maps showing the steepness of various places on the surface of the Earth. The angle $\theta$ is measured from the "north pole" of the black hole. The ring singularity is at $\theta = \pi$ at a distance $a = 0.6$ from the center of the black hole. The Boyer-Lindquist radial coordinate $r$ is zero at the ring singularity; ∞ at the center of the black hole, and goes over asymptotically to the conventional radial distance far from the black hole. Contrary to the impression one receives from the standard picture in Figure 1, curvature is *far* from constant, and is *negative* over large regions of this phase space, as noted by some workers [8]. A more direct comparison between the two pictures is possible if we "unpack" one of these invariants and plot it for several representative values of $r$ in spherical polar coordinates (Figure 3). The contrast could hardly be more dramatic!

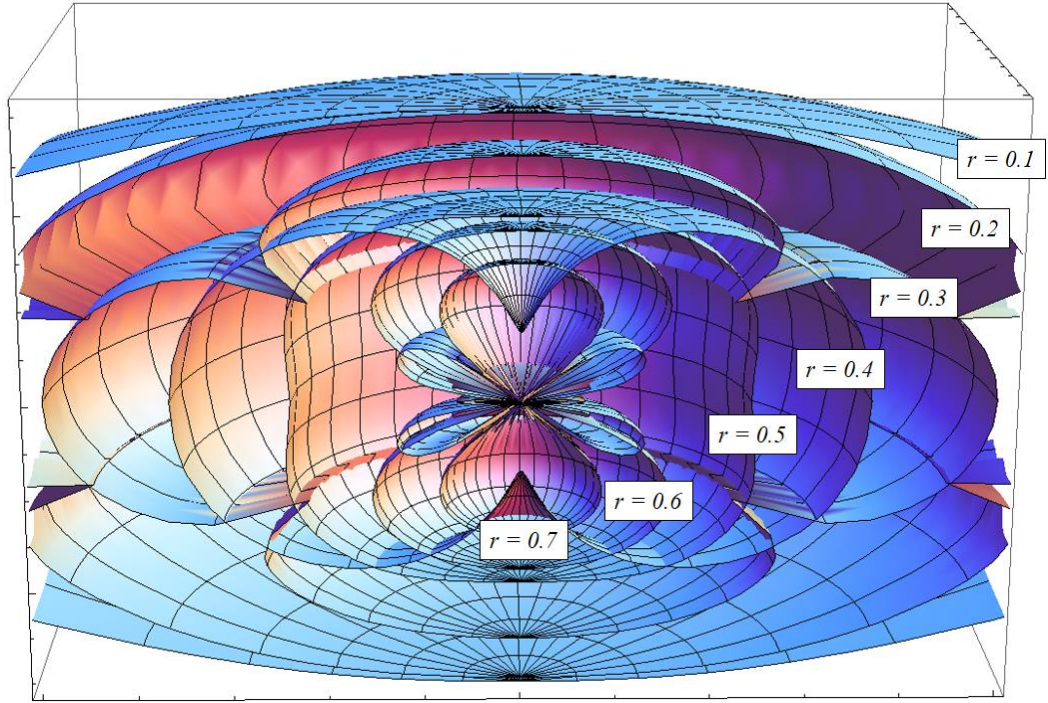

**Figure 3**: *The first Weyl invariant $I_1$, plotted in spherical polar coordinates for several representative values of r (compare Figure 1, right).*

Figures 2 and 3 show that the Weyl invariants gyrate wildly near the singularity; the mixed invariants less so. The pure Ricci invariant is well-behaved by comparison. This is interesting since the Weyl tensor encodes the degrees of freedom corresponding to a free gravitational field [9]. The crenellations themselves have been attributed to conflicting contributions to the curvature from the gravito-electric and gravito-magnetic components of this field, the latter generated by the black hole's rotation [10,11].

The importance of invariants in physics can hardly be overstated. The single invariant $I_5$ (also known as the Ricci curvature scalar) contains within itself the "seed" of General Relativity through its role as the action of that theory [2]. Beginning instead with a different action based on $I_1$ leads to an entirely different theory of gravity, conformal Weyl gravity [12]. The potential of the other invariants is as yet unexplored. We do not know whether the expressions we have found will eventually lead to greater physical insight into the nature of black holes. But we do know that the first step in obtaining any such insight is to *find* the invariants! And that is what we have done here.